\documentclass[trackchanges]{aastex7}

\usepackage[T1]{fontenc}
\usepackage{siunitx}
\usepackage{upgreek}
\usepackage{subcaption}
\usepackage{xcolor}
\usepackage{graphicx}	
\usepackage{amsmath}	
\usepackage{dirtytalk}

\begin{document}

\title{Observability of radio reflections from exoplanet ionospheres with next generation radio telescopes}

\author[0000-0001-9079-7794]{Bhavesh Jaiswal}
\affiliation{Space Astronomy Group, \\U R Rao Satellite Centre, \\Indian Space Research Organization, \\Bengaluru 560037, India}
\affiliation{Department of Physics, \\Indian Institute of Science, \\Bengaluru 560012, India}
\email[show]{bhavesh@ursc.gov.in}  

\author{Nirupam Roy}
\affiliation{Department of Physics, \\Indian Institute of Science, \\Bengaluru 560012, India}
\email{}

\begin{abstract}

Much has been learned about exoplanets and their atmospheres in the last three decades with the help of highly sensitive optical telescopes. Limited observations using X-ray telescopes have revealed the presence of ionospheres with very high density plasma around the hot Jupiter HD189733b. Owing to high density, the cutoff frequency of this plasma would lie in the range of few GHz. As the planet goes around the star, we suggest it might be possible to capture the stellar radio emission reflected from the ionosphere of the planet. We find that the reflected spectrum has a slope which is representative of the plasma density profile of the ionosphere and has a cutoff frequency. After investigating the reflection and free-free absorption process in the ionosphere, we find that this reflected signal, though feeble, can be captured by very sensitive radio telescopes operating in the low frequency range. We estimate the reflected signal from the ionosphere of a hot Jupiter and find that the flux ratio of the planet to the star are about $\sim 0.01\%$. In the view of development of facilities like Square Kilometer Array, it might be possible to capture the reflected radio signal from the ionosphere and constrain the thermal state of the ionosphere.

\end{abstract}



\section{Introduction} \label{sec:intro}

So far, more than 5000 exoplanets have been discovered around other stars (https://exoplanetarchive.ipac.caltech.edu/). Of these, the transiting exoplanets provide a unique opportunity to study the spectroscopic signatures of the atmosphere \citep{2010ARA&A..48..631S,2015PASP..127..941C}. Many of the ground based as well as space based telescope have revealed the presence of water, carbon-di-oxide, methane, sodium etc. in the atmospheres of transiting exoplanets\citep{2018haex.bookE.100K}. Much of what we have learned so far about exoplanets comes from the observations in the visible and IR bands. There have been very little observations in X-ray and radio bands. One main reason for the lack of X-ray and radio observations has been the sensitivity of telescopes to probe the star or the planet except in visible and IR bands. Despite limited observations the results have been very encouraging \citep{2018A&A...612A..52O,2021sf2a.conf..491G,2023NatAs...7..569P}. Recently, there has been an increased interest in low frequency radio observations to establish the presence of auroras on exoplanets. There have been several predictions to push for the radio observations of planet hosting stars \citep{2015aska.confE.120Z, 2015ASSL..411..213G}.

The sensitivity of radio observations is slated to improve drastically given the commencement of Square Kilometer Array (SKA) \citep{2009IEEEP..97.1482D,2019MNRAS.484..648P} and planned next generation of Very Large Array (ngVLA). Given the planned drastic improvement in the sensitivity of radio telescopes in future it is imperative to look for the possible signatures of exoplanets in radio band. One such recent attempt was made by \cite{2019MNRAS.484..648P} where possible signatures of radio transits were discussed. The radio transits were also found to cause unusually high transit depths when occulting an active region on the star \citep{2013ApJ...777L..34S}. The calculations by \citep{2020ApJ...895...62S} demonstrate the large flux contributed by the ionospheres of hot Jupiters in radio frequencies.

In this work, we discuss the radio waves reflected from the ionosphere of exoplanets. The reflection of radio waves from the ionospheric plasma is a well known phenomenon \cite{1986rpa..book.....R}. In fact, much of the radio transmission in Earth's atmosphere is due to the reflection of radio waves from the ionosphere. This allows the radio waves travel longer distances than otherwise. An ionospheric plasma density having N electrons $e^-$ /$\mathrm{ m^{3}}$  has a cutoff frequency ($\nu=f_c$) which is given by $f_c=9\sqrt{N}$ Hz. Any radiation with $\nu<f_c$ is reflected and the rest is transmitted. The Earth’s ionosphere has a plasma density of $10^{12} \mathrm{e^- /m^{3}}$ which makes the cutoff frequency to be about $\sim 10$ MHz.

There are few observations existing about the ionosphere of exoplanets. One notable effort by \cite{2013ApJ...773...62P} using Chandra X-ray Observatory, finds the ionosphere of hot Jupiter HD 189733b is extended much beyond the optical/IR radius of the planet and can have electron density as high as $7\times 10^{16} \mathrm{(e^- /m^{3})}$ at $1.75R_p$; where $R_p$ is the planetary radius measured during optical transits. Such high electron densities upto this height have no other analog in the solar system. This is likely caused by the extreme proximity of the planet to the star, which leads to unusually high flux of UV and X-ray on the planet and rips the electrons from their molecules. Given the high number density of plasma in the ionosphere of these hot Jupiters, it is expected that the high frequency radio waves are reflected from these ionosphere. For the case of HD 189733b this cutoff frequency would be close to 2 GHz. Radio emission originating at the star can get reflected from the planetary ionosphere if the cutoff frequency conditions are met. Interestingly, the future radio observatories, such as the Square Kilometer Array have a broadband coverage of $\sim$ 50 MHz to 30 GHz \citep{2009IEEEP..97.1482D}, which combined with its high sensitivity might make them suitable for detecting the reflected radio waves from the high density plasma of planets in close proximity to the star. 

The reflections of radio waves from the ionospheres of the solar system planets are not well documented mainly because the plasma density of the ionospheres is less than or similar to that of Earth. This unfortunate coincidence ensures $f_c (\mathrm{planets}) \lesssim f_c (\mathrm{Earth})$ and hence radio reflections from other solar system planets remain un-observable for telescopes within the Earth's ionosphere. However, space based observatories operating in sub-MHz frequencies can measure this reflected radiation. One example is the MARSIS instrument on Mars Express mission around Mars which observes the reflected waves from the ionosphere \citep{2007P&SS...55..864N}. In these observations one can clearly notice the part of the spectrum with $\nu<f_c$ is reflected from the ionosphere and that with $\nu>f_c$ is reflected from  the surface. Since MARSIS is a sounding instrument in orbit, a larger delay in the reflected signal signifies a larger distance of the reflecting surface. Since ionospheric plasma densities reduce exponentially with a certain scale height, it is expected that $f_c$ changes with altitude which can manifest as a relative delay for $\nu<f_c$.

The reflection from the ionosphere is considered to be specular in nature. While considering the reflection from the ionosphere it is important to consider the shape of the reflecting body. The ionospheres can be considered to be spherical in shape as they remain bound to the gravity of the planet. However the ionospheres can get disturbed due to the stellar activity. The hot-Jupiters which exist in close proximity of their host star ($<0.1$ AU) can be expected to get severely affected due to the high energy radiation and increased flux of stellar wind, especially at high altitudes where the effect of gravity is weak. Presently it is difficult to predict the shape of the hot-Jupiter ionospheres however we proceed with a set of reasonable assumption of spherical and ellipsoidal shapes. The consideration of ellipsoidal shape does not have any particular physical motivation but rather a first order attempt to consider a non-spherical shape. The vertical profile of plasma density in solar system planets is usually considered as a Chapman profile, which has a base altitude above which the number density decreases exponentially and below which it drops down drastically. In our calculations also we consider an exponentially decreasing plasma density with a scale height which is considered to be several 1000s of kilometers. In our heuristic approach, we ignore many second order effects such as plasma number density variation with longitudes, latitudes etc.
\begin{figure*}
\begin{center}
\includegraphics[width=0.9\textwidth]{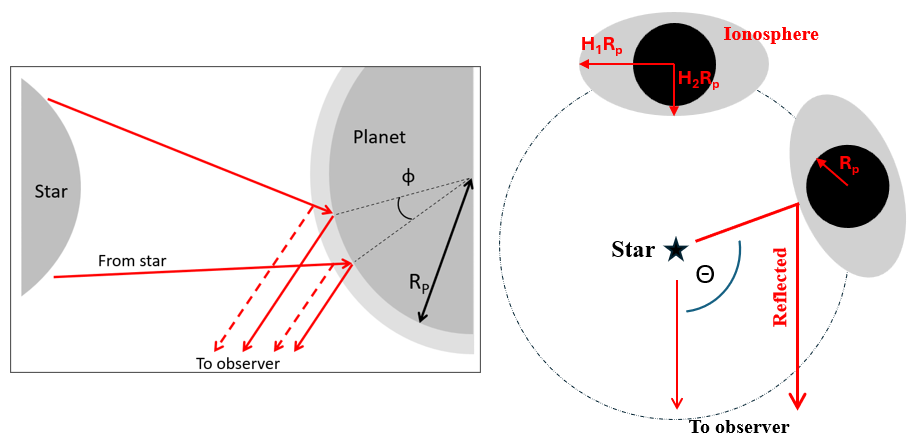}
\caption{Geometry of reflection from the planetary ionosphere. [Left] The stellar radio emission, reflected from two layers of spherical ionosphere. The angle $\phi$ shows the extent of reflection zone. The radiation reflected from the top layer -- shown with dashed lines has lower frequency than the radiation reflected from the lower layer which is shown with solid lines. [Right] The orbital geometry of planet-star system. The star is marked in the centre and the planetary ionosphere is shown in an ellipsoidal shape around the planet. At any moment, the rays originating at the star as well as the reflected rays are travelling towards the observer. Phase angle $\theta$ is also marked.}\label{geometry}
\end{center}
\end{figure*}

\section{Signal estimation}\label{sec:Res}
To study the radio reflections from exoplanets we proceed by considering HD189733b \citep{2005A&A...444L..15B} as a fiducial case. We assume a fully ionized ionosphere at all levels above the base altitude. We adopt the plasma density values from \cite{2013ApJ...773...62P} and assume that the atmosphere is fully ionized. We consider an inclination angle of $\ang{90}$ such that the planet is fully illuminated during the eclipse conditions. For brevity, we refer the superior conjunction or the secondary eclipse of the planet as just eclipse here. We are not interested in the transit of the planet as this would not lead to any reflection from the ionosphere although it may lead to scintillation and lensing of the star \citep{2019MNRAS.484..648P}. The radio waves are considered to be reflected in an specular manner from the ionosphere which is considered to be smooth at a scale of the wavelength of the radiation. The following Figure \ref{geometry} shows the geometry of reflection from the exoplanet. 

We proceed with doing a first order calculation of the reflected signal. Considering the stellar intensity as $I$, the stellar radius as $R_S$ ,the planetary radius (at ionospheric altitudes) as $R_P$, the planet-star separation as $a$, and the distance to this planetary system as $d$, one can calculate the extent of the \textit{reflection zone} (illuminated region causing specular reflection in the direction of observer) subtends an angle $\phi \approx R_S/a$ at the planet centre, assuming $a\gg R_S$. This reflection zone having a linear size of $\approx R_P\phi$ on the planet, subtends a solid angle $\Omega$ at the observer given by,

\begin{equation}\label{eq_signal1}
\Omega = \pi\left(\frac{R_PR_S}{ad}\right)^2.
\end{equation}

The stellar flux reflected from the planetary ionosphere in the direction of the observer is then given by

\begin{equation}\label{eq_signal2}
F_P = I\pi\left(\frac{R_PR_S}{ad}\right)^2,
\end{equation}

and the stellar flux at the location of the observer is then given by 

\begin{equation}\label{eq_signal3}
F_S = I\pi\left(\frac{R_S}{d}\right)^2.
\end{equation}

The ratio of the ionospheric reflection and the stellar flux is then given by:

\begin{equation}\label{eq_signal4}
\frac{F_P}{F_S} = \left(\frac{R_P}{a}\right)^2.
\end{equation}

\begin{figure*}
\begin{center}
\includegraphics[width=1\textwidth]{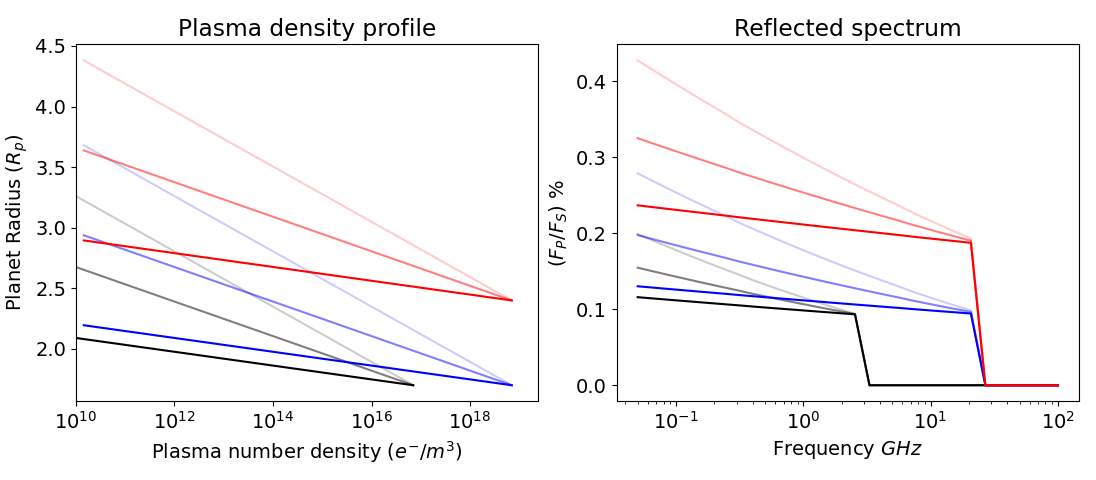}
\caption{[Left] Ionosphere profiles are marked for various cases with lines of different transparencies. The scale heights of 2000 km, 5000 km and 8000 km are marked with increasing transparencies. The black and blue profiles have their ionosphere base altitudes at 1.7$R_P$ and the red profiles have their ionosphere base altitude at 2.4$R_P$. The plasma density is considered $0$ below the ionosphere base altitude. [Right] Reflected spectrum, without considering free-free absorption, corresponding to the profiles on the left panel. The color coding is the same as that on the left panel.}\label{spectrum}
\end{center}
\end{figure*}

It is noteworthy here that the reflected signal from the ionosphere comes from a region of the ionosphere which subtends a 'maximum' cone angle angle $\phi$ at the centre of the planet, where $\phi = R_S/a$. The calculations presented here are similar to that of \cite{doi:10.1089/ast.2022.0101} and the reflectivity (albedo) of the ionosphere is assumed to be 1. For the case of HD189733b \citep{2013ApJ...773...62P} we assume that $R_P=1.75\times 1.13\times R_J$, where $R_J$= Jupiter radius. With these numbers one can estimate, to first order, that $(F_P / F_S)\approx 0.1\%$.

In order to do the detailed calculations of reflection across different phase angles we create a simplistic framework of ray-tracing. Here, we send light rays from the direction of the observer across the illuminated part of the planet and count the number of rays which are reflected in the direction of the star and intercept the stellar disc. We consider a specular reflection and divide the planetary disc in a $400\times 400$ grid on which reflection is considered. The light rays reflected from a region within the reflection zone will be reflected in the direction of the star. The fraction of the light rays intercepted by the stellar disc to that of the total number of rays provides the fraction of the planet to star flux. 

\begin{figure*}
\begin{center}
\includegraphics[width=0.5\textwidth]{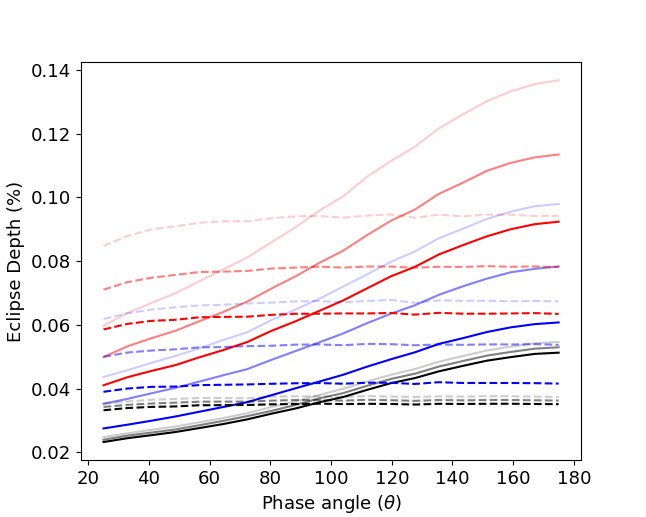}
\caption{The reflected light from the planet as a function of phase angle $\theta$. The colors and the transparencies of the lines correspond to Figure \ref{spectrum}. The solid lines correspond to the ellipsoidal ionosphere and the dashed lines correspond to the spherical ionospheres. For the ellipsoidal ionospheres $H_1=1.7$ and $H_2=1.3$ whereas for the spherical ionospheres $H_1=1.7$ and $H_2=1.7$.}\label{light_curve}
\end{center}
\end{figure*}

Along with the fiducial profiles of HD189733b we consider various kinds of plasma density profiles of the ionospheres. The various cases of plasma density profiles are shown in the left panel of Figure \ref{spectrum}. The plasma density $N$ is seen to be increasing as we go deeper into the atmosphere, up-to a base altitude below which it is considered to be zero. In these simulations we consider a near-eclipse geometry i.e. $\theta \approx \ang{160}$. As discussed above, $f_c=9\sqrt{N}$ and hence the smaller frequencies are reflected from the upper parts of the atmosphere. Due to this the planet appears to be larger at lower frequencies and consequently has a larger reflected signal, see Eq. \ref{eq_signal4}. The reflected spectrum of the specularly reflected radio waves is shown on the right panel of Figure \ref{spectrum}. The higher frequencies are reflected from deeper parts of the ionosphere. In this simplistic approach we do not consider any refraction in the ionosphere. The highest frequency reflected corresponds to the plasma density at the base of the ionosphere. This would also mark a sharp cutoff in the reflected spectrum beyond which no frequency is reflected (and is considered to be absorbed in deeper layers of atmosphere). This sharp cutoff could be a discernible feature in a broadband radio spectrum.

\begin{figure*}
\begin{center}
\includegraphics[width=1\textwidth]{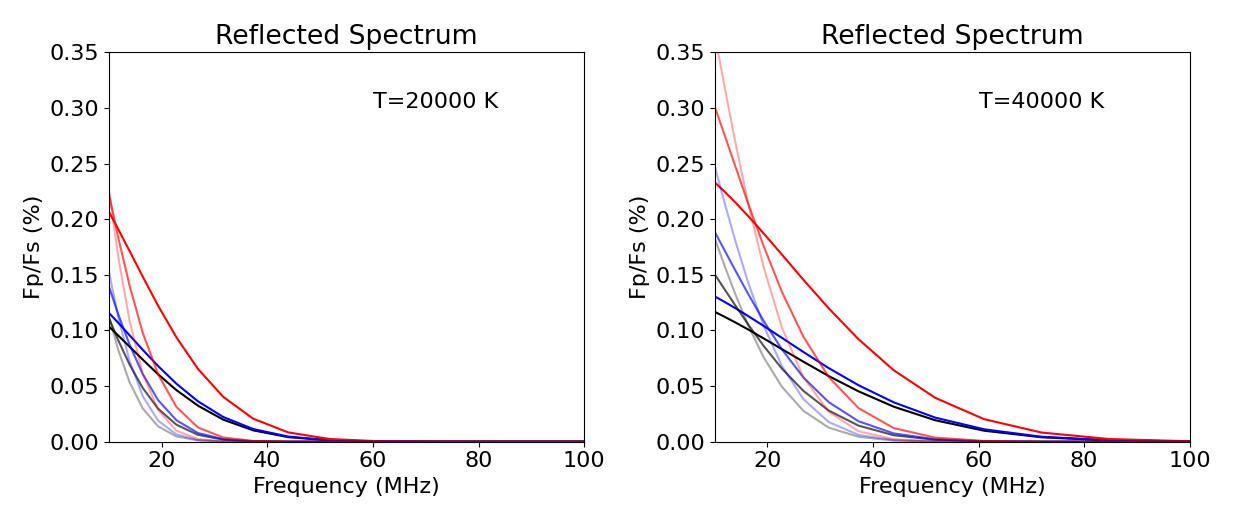}
\caption{[Left] Reflected spectrum, considering free-free absorption, corresponding to the profiles in Figure \ref{spectrum}, are shown for an ionospheric temperature of 20000 K [middle] and 40000 K [right] for the all the profiles. The color coding is the same as that on the left panel. The ionospheric scale height and the base altitudes are same as that in Figure \ref{spectrum}.}.\label{spectrum2}
\end{center}
\end{figure*}

So far, we have discussed the reflected signal considering only the reflection properties of the ionospheric plasma, however the radiation going through the plasma will also undergo free-free absorption. The free-free absorption optical depth \citep{2016era..book.....C} at a particular frequency ($\nu$) is dependent on the Emission Measure (EM) and ionospheric temperatures (T) in the following manner:

\begin{equation}\label{eq_free-free opt dep}
\tau = 3.28\times10^{-7} (\frac{T}{10^4})^{-1.35}\nu^{-2.1}EM.
\end{equation}
Here, the temperature is in Kelvin and the frequency is in GHz, and the EM in the units of $pc/cm^6$ is defined by integrating the square of plasma density ($n$) along the line of sight ($ds$) in the following manner:

\begin{equation}\label{eq_free-free EM}
EM = \int n^2 ds.
\end{equation}
Given the large ionospheric path lengths and low temperatures, this absorption can be significant and can significantly reduce the strength of the reflected signal. We estimate the free-free absorption optical depths in our models for each frequency and consider the reflection only if $\nu \leq f_c$. It is noteworthy here that the free-free absorption is dependent on the ionospheric temperatures too (as per Eq. \ref{eq_free-free opt dep}) which we consider to be vertically uniform in our models. Based on \cite{2013ApJ...773...62P} we assumed two cases of the ionospheric temperatures -- a realistic scenario of 20000 K and an optimistic scenario of 40000 K. We find that in this scenario, only the lower frequencies are not catastrophically absorbed. The higher frequencies, which require a larger $f_c$ have to travel deeper into the atmosphere, which, due to exponential plasma density profile considered in our models, renders a longer path length and higher values of $\tau$ eventually leading to a complete absorption. In figure \ref{spectrum2} we show the reflected spectrum for the two cases of ionospheric temperatures. The smaller scale height, will reduce the values of $\tau$ for higher frequencies and hence can allow reflection from higher frequencies as compared to the larger scale heights. Also, increasing the ionospheric temperature helps to reduce the overall free-free absorption and is favorable for studying reflections. In our simulations, we did not find any significant reflection component beyond 100 MHz frequency range, which seriously brings down the frequency limit for detecting reflections, however still makes it an interesting case for SKA-low which will operate at frequencies 50 MHz or higher. Due to the free-free absorption, the effect of cut-off frequency is not seen as a step function but rather as a gradual slope in the reflected component. We also observe that a reflection of $\sim 0.01$ to $0.05\%$ is observed at the 50 MHz band (low frequency edge of SKA) in only a few plasma profile of 40000 K case and all other cases have much lower reflection.

As the planet orbits the star, it would be possible to sample the spectrum at different values of $\theta$, to produce a phase curve. In the optical/IR bands this is well observed, especially for hot-Jupiter planets (such as in \cite{2012A&A...548A.128D}) where the observed intensity of star+planet system increases as the star eclipses the planet. This happens mainly because the hot-spot of the planet lies at the sub-stellar point and comes directly in the view of the observer for $\theta$ close to 0. Our case however, is different as we consider a specular reflection from the ionosphere. Since the entire ionosphere is reflective, there is no change in the reflected flux with $\theta$ leading to a constant flux in the phase curve. The phase curve, however, can show variation if the shape of the ionosphere is not strictly spherical. This can happen due to the ram pressure of the radiation and stellar wind on the planets, particularly those which lie in close proximity of the star, such as hot-Jupiters. We assume a spheroidal ionosphere and calculate the phase curve. The phase curve for various cases of plasma density are shown in Figure \ref{light_curve} for the spherical and ellipsoidal case. Ellipsoidal shape, having a larger projected area offers more reflected flux during the eclipses (i.e. $\theta \sim \ang{180}$).

It must be noted here that the hot Jupiter ionospheres are much brighter in the radio band \citep{2020ApJ...895...62S}. With a simplistic model they have shown that the thermal emission from the ionosphere can be about few 10s of percentage of the stellar emission and it has a frequency dependence. The reflected signal however, as discussed here, will be observed over and above the emission signal. 
A detailed analysis of radio signal from the ionosphere would require the layer-by-layer calculation of emission added with the reflection part considering the plasma cut-off frequency for each layer. In a subsequent paper we will assess the feasibility of radio reflection added with the radio emission part in the range of SKA-1 and SKA-2 and asses its feasibility with SKA-low.

\section{Observability with future telescopes} \label{sec:forw_scat}

Due to close proximity of the planet to the star, it would not be possible to observe the planet alone, rather planet and the star as an integrated object. The stellar radiation in the radio band can be a combination of the thermal as well as non-thermal processes. In order to make the observations of the reflected radio radiation, one needs to rely on bright radio sources. In order to detect faint reflections, of the order of $\sim 0.01\% - 0.05\%$, it is essential that source brightness is at least about $\sim10000 - 2000$ times the noise limit of the telescope for a particular integration time. If this condition is not met, one needs to rely on longer integration times or combining many observations in order to improve the sensitivity. 

A comparison of broadband sensitivity of the present and future radio telescopes in the context of exoplanets is discussed in Figure 1 of \cite{2019MNRAS.484..648P}. The sensitivity limit of Square Kilometer Array; SKA2 at about 50 MHz and for 1 hour integration time lies at 20 $\mu \mathrm{Jy}$ per beam respectively. As per Figure \ref{spectrum2}, the detection of radio reflections with an SNR of $\sim3$ from SKA2 for $\sim 0.01\% - 0.05\%$ reflection, would demand the radio brightness of the star to be about $\sim200$ mJy and $\sim40$ mJy respectively for a $\sim 9$ hour integration time. 

The solar emissions in radio, during active phase, can range from $10^7$ Jy to a maximum of $10^{10}$ Jy during extreme radio flares (see Fig. 2 of \citep{2012P&SS...74..156Z}). For a Sun-like star located at 2 parsecs, this would translate to 0.6 to 60 mJy flux levels. Further, a recent catalogue of radio stars \citep{2024arXiv240407418D} has measured the radio intensities of about 800 stars ranging from ultracool dwarfs to Wolf-Rayet stars. The published results show that the peak radio fluxes can have a large range and are observed to be in the range of 0.1 mJy to <100 mJy (upper limit), with peak flux rarely going beyond 10 mJy levels. Though there is no mention of the presence of planets around these stars but if the trend of radio fluxes is similar around other planet hosting stars we are hopeful of detecting the ionospheric reflections in some of the bright and stable system, though we note the caveat that the kind of the integration times required would increase appropriately with decreasing brightness of the star. These observations can benefit from a radio stellar monitoring program which may allow the observer to observe during the bright period.  These estimate point towards much longer integration times, several 10s (and possibly 100) of hours for the faint sources in order to detect a reflected signal. For such long observation times the stability of the stellar radiation need to be characterized before the observation, since many of the bright radio emissions are known to come out as bursts rather than steady emissions. It is noteworthy here that since we are attempting to detect the reflected flux in the total flux of the star-planet system, the distance of the star from Earth does not matter as long as the flux levels are high enough.

The main limitation, as we see, in detecting the faint effects of reflection will come from the stellar flux variability in the radio band during the observations. The quiet Sun itself is known to have variations of over few percentage in the radio flux over a timescale of minutes to hours apart from the 27 day period variation which can be about few 10s of percentage (for example, see figure 1 of \cite{2018MNRAS.475.3117B}). Since the eclipse duration can be about few hours, our main concern would be about the hourly variations. The radio reflections can be corrupted by stellar flares and even the signature of Star-Planet Interactions. Without going into the details, we suggest two possible ways to mitigate or minimize such corruptions in the signal. Firstly, the observations should preferably be done at large phase angles i.e. when the planet is nearing the eclipse position. This would allow the reflected signal to maximize and simultaneously minimizing the star-planet interaction signatures which is usually highest at the quadrature angles \citep{2023MNRAS.524.6267K}. Secondly, to avoid the signal corruption by flares, we need to look for the signatures of spectral slope as well as the cut-off frequency in the observations (as shown in Figure \ref{spectrum2}). Also, co-adding several observations over many orbits of the planet would help to average out any such affects, especially when the in and out of eclipse observations are compared considering that the spectral nature of the stellar contribution and the reflected part will be different. A relevant discussion for disentangling stellar variability from these observations can be found in \citep{2019MNRAS.484..648P} where some practical strategies like simultaneous multi-observatory observations, such as SKA and JWST are suggested. A radio-survey of nearby planet hosting stars will definitely help in selecting a target which is comparatively inactive in radio bands and hence easier to observe.

In the simulations here, we consider the reflection from a smooth spheroidal ionosphere however in a realistic situation the shape of the ionosphere can have non-uniformities on smaller scales though the planetary scale ionospheric structure can still have a near-spheroidal shape. Such small scale non-uniformities can lead to a spread in the area of the reflection zone ($\theta$) which is same as the angular extent of the star. This effect would be similar to the reflection of sunlight from a rough ocean surface where the small scale waves on the water surface can make the reflection zone much larger. However, increase or decrease in the extent of the reflection zone does not increase or decrease the total reflected intensity as long as the non-uniformities are much smaller than planetary scale.

\section{Discussion and Conclusions}\label{sec:Disc}
The ionospheres of the exoplanet have remained largely elusive and the radio astronomy might offer a view on the dense ionospheres of highly illuminated hot Jupiters by observing the radiation reflected from the ionospheres. Coincidentally, the cutoff frequencies of these dense ionospheres can be within the spectral limits of the ground based telescopes. With more advancement planned in terms of sensitivities of the Square Kilometer Array we are hopeful of detecting the radio reflections from the ionospheres of hot-Jupiters. 

We show that the reflected spectrum can have a slope which can be representative of the the scale heights of the plasma in the ionosphere. It might give clues to the scale heights of plasma density profile which can help us to understand the thermal state of the ionosphere. We also note that when considering a realistic scenario and free-free absorption along the ray line of sight, most of the high frequency radiation is getting absorbed before the reflection and such broad frequency slopes seem to be impossible to be detected in reflection. However, the reflection at the lower frequencies can still be observable, more like a step function increase (as shown in Figure \ref{spectrum2}) rather than a gradual increase. Along with the scale height, the features like the ionospheric base densities can be captured via the abrupt spectral cutoff in the reflected spectrum.

We recommend the cataloging of radio emissions from the planet hosting stars and also measuring their short term variability. This exercise will help us in target selection.  Also, we suggest more X-ray transit observations of other hot Jupiters which will allow us to understand the ionosphere of these planets. We note that the short term variability due to flares or radio bursts can make these observations challenging and extra care must be taken to avoid such events during the observations. Though we consider a non-inclined planet in our simulation, the reflection can happen from a inclined planet as well (such as by the majority of the planets discovered via Radial Velocity method).

\begin{acknowledgments}
\textbf{Acknowledgement:} We acknowledge the useful comments by Prof. Benjamin Pope which helped us in improving the quality of this manuscript. We acknowledge the comments of the anonymous reviewers and their careful look at the calculations which has helped us improve the results and discussion.
\end{acknowledgments}


\bibliography{bibliography}{}
\bibliographystyle{aasjournal}



\end{document}